\documentclass[reprint,aps,prb]{revtex4-2}
\usepackage{dcolumn}
\usepackage{bm}%
\usepackage{graphicx,graphics,color,epsfig}
\usepackage{multirow,booktabs}
\usepackage{float}
\usepackage[normalem]{ulem}
\usepackage{xcolor}
\usepackage{amsmath}
\usepackage{amssymb}
\newcommand\ii{\mathrm{i}}
\newcommand\ee{\mathrm{e}}
\newcommand\dd{\mathrm{d}}
\newcommand\nn{\nonumber}
\usepackage{hyperref}

\begin{document}

\title{Dynamical zero modes and boundary dependence in dynamical quantum phase transitions}
\author{Siyan Lin}
\affiliation{Beijing National Laboratory for Condensed Matter Physics, Institute of Physics, Chinese Academy of Sciences, Beijing 100190, China}
\affiliation{School of Physical Sciences, University of Chinese Academy of Sciences, Beijing 100049, China}

\author{Xu Feng}
\affiliation{Beijing National Laboratory for Condensed Matter Physics, Institute of Physics, Chinese Academy of Sciences, Beijing 100190, China}
\affiliation{School of Physical Sciences, University of Chinese Academy of Sciences, Beijing 100049, China}

\author{Xiuhua Tian}
\affiliation{Beijing National Laboratory for Condensed Matter Physics, Institute of Physics, Chinese Academy of Sciences, Beijing 100190, China}
\affiliation{School of Physical Sciences, University of Chinese Academy of Sciences, Beijing 100049, China}

\author{Shu Chen}
\thanks{Corresponding author: schen@iphy.ac.cn}
\affiliation{Beijing National Laboratory for Condensed Matter Physics, Institute of Physics, Chinese Academy of Sciences, Beijing 100190, China}
\affiliation{School of Physical Sciences, University of Chinese Academy of Sciences, Beijing 100049, China}

\date{\today}

\begin{abstract}
    Boundary conditions are usually expected to cause only finite-size corrections to bulk quantities, but this expectation can fail for dynamical quantum phase transitions. In this work, we show that such boundary dependence is encoded in dynamical zero modes (DZMs) of the Loschmidt matrix, which are defined as singular vectors whose singular values vanish in the thermodynamic limit. Using the Su-Schrieffer-Heeger (SSH), extended SSH, and Kitaev models as examples, we find that the time interval where the Loschmidt rate functions (LRFs) under periodic and open boundary conditions differ coincides with the emergence of DZMs in the open-boundary Loschmidt matrix. These modes carry the boundary-dependent contribution: removing them from the open-boundary LRF recovers the periodic-boundary result. We further show that these DZMs lead to finite-precision numerical instability, since their finite-size singular values decay exponentially with system size and eventually become unresolved in fixed-precision arithmetic. A reliable small-size branch before this loss of precision can be used to estimate the LRF in the thermodynamic limit by linear extrapolation. Our results identify DZMs as both a diagnostic of boundary-dependent LRFs and the origin of the associated numerical instability.
\end{abstract}

\maketitle

\section{Introduction}

Understanding nonequilibrium quantum dynamics is a central problem in modern many-body physics~\cite{RevModPhys.83.863,QuenchDynamics}. A particularly useful concept is the dynamical quantum phase transition (DQPT), which describes nonanalytic behavior during real-time evolution after a quantum quench \cite{PhysRevLett.110.135704,Heyl_2018,Heyl_2019}. In a typical quench protocol, a system is prepared in an initial state $|\Psi_0\rangle$, often the ground state of an initial Hamiltonian, and is subsequently evolved under a different Hamiltonian $H_f$. The central quantity is the Loschmidt amplitude
\begin{equation}
    \mathcal{G}(t)=\langle \Psi_0|\ee^{-\ii H_f t}|\Psi_0\rangle ,
\end{equation}
which is the return amplitude of the time-evolved state to the initial state. For a many-body system, the Loschmidt echo $|\mathcal{G}(t)|^2$ usually takes a large-deviation form, motivating the definition of the Loschmidt rate function (LRF)
\begin{equation}
r(t)=-\lim_{L\to\infty}\frac{1}{L}\ln\left(\left|\mathcal{G}(t)\right|^2\right) .
\end{equation}
In analogy with the partition function and free-energy density in equilibrium statistical mechanics, $\mathcal{G}(t)$ and $r(t)$ play the roles of a dynamical partition function and a dynamical free-energy density, respectively. DQPTs are then identified by nonanalyticities of $r(t)$ at critical times. Since their introduction, DQPTs have been extensively studied in spin models \cite{PhysRevB.89.161105,PhysRevB.89.125120,PhysRevB.97.134306,PhysRevLett.123.160603,PhysRevB.102.094302,PhysRevB.104.094311,PhysRevX.11.041018,PhysRevResearch.4.033032,PhysRevA.110.042209,srx7-cpl4,qc3w-s38g,sbdw-2ly3,wsxd-k6dv}, topological systems \cite{PhysRevB.91.155127,PhysRevLett.117.086802,PhysRevB.97.064304,PhysRevB.107.134302,PhysRevB.110.054312,PhysRevB.110.224302}, long-range interacting systems \cite{PhysRevB.96.134427,PhysRevE.96.062118,PhysRevB.96.125113,PhysRevLett.120.130601,PhysRevB.106.024311,PhysRevLett.130.100402}, disordered fermion systems \cite{PhysRevB.95.184201,PhysRevA.97.033624,PhysRevResearch.5.033178,PhysRevA.109.043319,PhysRevA.111.042208}, and open quantum systems \cite{PhysRevB.97.045147,PhysRevB.98.134310,PhysRevLett.125.143602,vvf8-ktpq,gqv9-rfss}.

By analogy with conventional thermal phase transitions, where boundary conditions usually do not affect the bulk free-energy density of short-range systems in the thermodynamic limit, one might expect the LRF to be independent of boundary conditions. Recent studies, however, have shown that this expectation is not generally valid for DQPTs. In Refs. \cite{PhysRevLett.123.160603,sbdw-2ly3}, the authors analyzed the transverse field Ising model using the renormalization group technique, and found that boundary conditions can become relevant for DQPTs at the unphysical fixed point. These results show that the boundary dependence of LRFs is a genuine thermodynamic effect rather than a finite-size artifact.

The above observations raise a natural question: 
What is the microscopic mechanism underlying the boundary sensitivity of DQPTs, and how is it encoded in the Loschmidt amplitude? In this work, we address this question from the viewpoint of the Loschmidt matrix. For free fermions, the many-body Loschmidt amplitude can be expressed as the determinant of a Loschmidt matrix constructed from correlation matrices \cite{PhysRevA.75.032333,PhysRevB.97.064304}. The singular values of this matrix then provide a direct way to resolve how different parts of the spectrum contribute to the LRF. In a finite system, the relevant singular values are generally nonzero. We refer to singular values that decay exponentially with system size and eventually vanish in the thermodynamic limit as thermodynamic zero singular values (TZSVs). Although their number remains finite, they can make a finite contribution to the LRF after the logarithm is taken and divided by the system size. We refer to the singular vectors associated with TZSVs as dynamical zero modes (DZMs). Thus, TZSVs denote the singular values, whereas DZMs denote their associated singular vectors. Such DZMs can therefore serve as a diagnostic of the time interval in which LRFs do not coincide under different boundary conditions.

This DZM viewpoint also exposes a closely related numerical problem. At finite size, DZMs appear as singular vectors with exponentially small but nonzero singular values. As the system size increases, these singular values eventually fall below the resolution of fixed-precision arithmetic. Once this happens, their contribution to the LRF can no longer be evaluated reliably, and the numerically computed LRF deviates from the correct result. Thus the DZMs play two roles: they encode the boundary-dependent part of the LRF, and they are responsible for finite-precision numerical instability.

The rest of the paper is organized as follows. In Sec. \ref{sec:exact_benchmarks} we focus on the Su-Schrieffer-Heeger (SSH) model as an illustrative example, and present exact benchmarks for the LRFs under periodic (PBC) and open (OBC) boundary conditions for the fully dimerized quench. In Sec. \ref{sec:singular_values} we identify DZMs of the open-boundary Loschmidt matrix and show how their TZSVs diagnose the interval where the two LRFs differ. We then discuss the finite-precision instability caused by these DZMs. In Sec. \ref{sec:other} we extend the analysis to more general quench protocols in the SSH, extended SSH, and Kitaev models. Section \ref{sec:conclusion} summarizes our conclusions and discusses possible directions for future work.

\section{Exact benchmarks for boundary sensitive dynamics and numerical instability}
\label{sec:exact_benchmarks}

We begin with an exactly solvable quench in the SSH chain \cite{PhysRevLett.42.1698,PhysRevB.22.2099}, which provides a benchmark for comparing LRFs under PBC and OBC and for diagnosing numerical instability. We consider the SSH chain with $2L$ sites, where $L$ is the number of unit cells. The Hamiltonian is
\begin{align}
    H(\delta)=-J\sum_j{\left[\left(1+(-1)^j\delta \right)c_j^\dagger c_{j+1}+\text{H.c.}\right]},
\end{align}
where $J$ is the nearest-neighbor hopping amplitude, $\delta$ with $|\delta|\leq 1$ is the dimerization parameter, $c_j$ is the fermionic annihilation operator on site $j$, and H.c. represents the Hermitian conjugate. The system is topologically nontrivial for $0<\delta\le1$, while it is topologically trivial for $-1\le\delta<0$. For PBC, we impose $c_{2L+1}=c_1$ and sum over $j=1,2,\cdots,2L$. For OBC, the sum terminates at $j=2L-1$. Throughout this work, we consider the half-filled case, so the number of fermions is $L$.

We first study the quench from $\delta_i=-1$ to $\delta_f=1$, for which exact expressions for LRFs can be derived under both PBC and OBC. For the PBC case, we can perform the Fourier transform
\begin{align}
    c_{2j-1}&=\frac{1}{\sqrt{L}}\sum_k{\ee^{\ii kj}c_{k,A}},\nn\\
    c_{2j}&=\frac{1}{\sqrt{L}}\sum_k{\ee^{\ii kj}c_{k,B}},
\end{align}
where we have introduced $A$ and $B$ as sublattice indices. Then the Hamiltonian can be written in the standard two-band form
\begin{align}
    H(\delta)=\sum_k{
    \begin{pmatrix}
        c_{k,A}^\dagger & c_{k,B}^\dagger
    \end{pmatrix}
    \boldsymbol{d}_k(\delta)\cdot\boldsymbol{\sigma}
    \begin{pmatrix}
        c_{k,A} \\ c_{k,B}
    \end{pmatrix}},
    \label{eq:two-band}
\end{align}
with
\begin{align}
    \boldsymbol{d}_k(\delta)=\left(-J(1-\delta)-J(1+\delta)\cos{k},-J(1+\delta)\sin{k},0\right).
\end{align}
Here $\boldsymbol{\sigma}=(\sigma_x,\sigma_y,\sigma_z)$ are Pauli matrices.

For a two-band model, the Loschmidt amplitude is given by \cite{PhysRevB.91.155127,PhysRevB.93.085416}
\begin{align}
    \mathcal{G}_{\text{PBC}}(t)=\prod_k\left[\cos(\epsilon_k^f t)+\ii\hat{\boldsymbol{d}}_k^i\cdot\hat{\boldsymbol{d}}_k^f\sin(\epsilon_k^f t)\right],
\end{align}
where $\epsilon_k^{i,f}=\sqrt{\boldsymbol{d}_k^{i,f}\cdot\boldsymbol{d}_k^{i,f}}$ and $\hat{\boldsymbol{d}}_k^{i,f}=\boldsymbol{d}_k^{i,f}/\epsilon_k^{i,f}$. The corresponding LRF is given by
\begin{align}
    r_{\text{PBC}}(t)
    &=-\frac{1}{L}\ln\left(\left|\mathcal{G}_{\text{PBC}}(t)\right|^2\right)\nn\\
    &=-\frac{1}{L}\sum_k{\ln\left[\cos^2(\epsilon_k^ft)+(\hat{\boldsymbol{d}}_k^i\cdot\hat{\boldsymbol{d}}_k^f)^2\sin^2(\epsilon_k^f t)\right]}.
\end{align}
In the thermodynamic limit, this becomes
\begin{align}
    r_{\text{PBC}}(t)=-\int_{-\pi}^\pi{\frac{\dd k}{2\pi}\ln\left[\cos^2(\epsilon_k^ft)+(\hat{\boldsymbol{d}}_k^i\cdot\hat{\boldsymbol{d}}_k^f)^2\sin^2(\epsilon_k^f t)\right]}.
    \label{eq:r_PBC1}
\end{align}
For the quench from $\delta_i=-1$ to $\delta_f=1$, we have
\begin{align}
    \epsilon_k^f=2J,\qquad \hat{\boldsymbol{d}}_k^i\cdot\hat{\boldsymbol{d}}_k^f=\cos{k},
\end{align}
and therefore
\begin{align}
    r_{\text{PBC}}(t)&=-\int_{-\pi}^\pi\frac{\dd k}{2\pi}\ln\left[\cos^2(2Jt)+\cos^2k\sin^2(2Jt)\right]\nn\\
    &=-2\ln\left(\frac{1+\left|\cos(2Jt)\right|}{2}\right)\nn\\
    &=-2\ln\max\{\cos^2(Jt),\sin^2(Jt)\}.
    \label{eq:r_PBC}
\end{align}

For the OBC case, the Loschmidt amplitude can be computed from a Slater determinant \cite{doi:10.1021/acs.jctc.5b01148,PhysRevResearch.5.033178}
\begin{align}
    \mathcal{G}(t)=\det{M(t)},
\end{align}
where $M$ is the Loschmidt matrix \cite{PhysRevB.97.064304} and its entries are given by
\begin{align}
    M_{ij}=\langle\phi_i|\ee^{-\ii H(\delta_f) t}|\phi_j\rangle,
\end{align}
with $|\phi_i\rangle$ being the $i$th occupied state. For the initial Hamiltonian with $\delta_i=-1$, the occupied states are
\begin{align}
    |\phi_j\rangle=\frac{1}{\sqrt{2}}\left(|2j-1\rangle+|2j\rangle\right),\quad j=1,2,\cdots,L.
\end{align}
The final Hamiltonian with $\delta_f=1$ consists of decoupled dimers on $(2,3),(4,5),\cdots,(2L-2,2L-1)$, together with two isolated edge sites 1 and $2L$. Using this structure, one obtains the following tridiagonal Loschmidt matrix:
\begin{align}
    M_L(t)
    =
    \begin{pmatrix}
        a&b&0&0&\cdots&0\\
        b&c&b&0&\cdots&0\\
        0&b&c&b&\cdots&0\\
        0&0&b&c&\ddots&0\\
        \vdots&\vdots&\vdots&\ddots&\ddots&b\\
        0&0&0&0&b&a
    \end{pmatrix},
    \label{eq:Loschmidt_mat}
\end{align}
where
\begin{align}
    a=\frac{1+\cos(2Jt)}{2},
    \quad
    b=\frac{\ii\sin(2Jt)}{2},
    \quad
    c=\cos(2Jt).
    \label{eq:abc}
\end{align}
The determinant of this matrix can be derived analytically, and the result is extremely simple,
\begin{align}
    \mathcal{G}_{\text{OBC}}(t)=\det{M_L(t)}=[\cos(Jt)]^{2L-2}.
    \label{eq:G_OBC}
\end{align}
The derivations of Eqs. \eqref{eq:Loschmidt_mat} and \eqref{eq:G_OBC} are given in Appendix \ref{app1}. Therefore, the LRF is
\begin{align}
    r_{\text{OBC}}(t)=-\frac{2L-2}{L}\ln{\cos^2(Jt)}.
\end{align}
In the thermodynamic limit, this becomes
\begin{align}
    r_{\text{OBC}}(t)=-2\ln{\cos^2(Jt)}.
    \label{eq:r_OBC}
\end{align}

\begin{figure}
    \centering
    \includegraphics[width=0.8\linewidth]{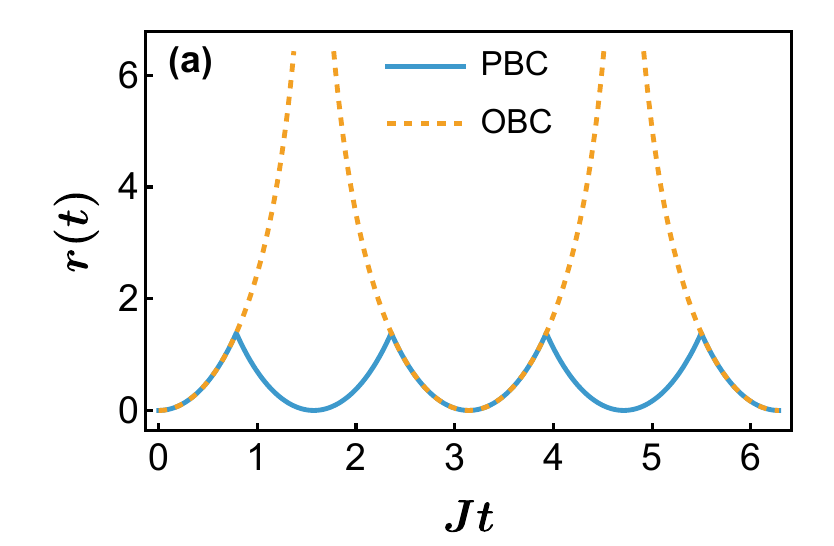}
    \includegraphics[width=0.8\linewidth]{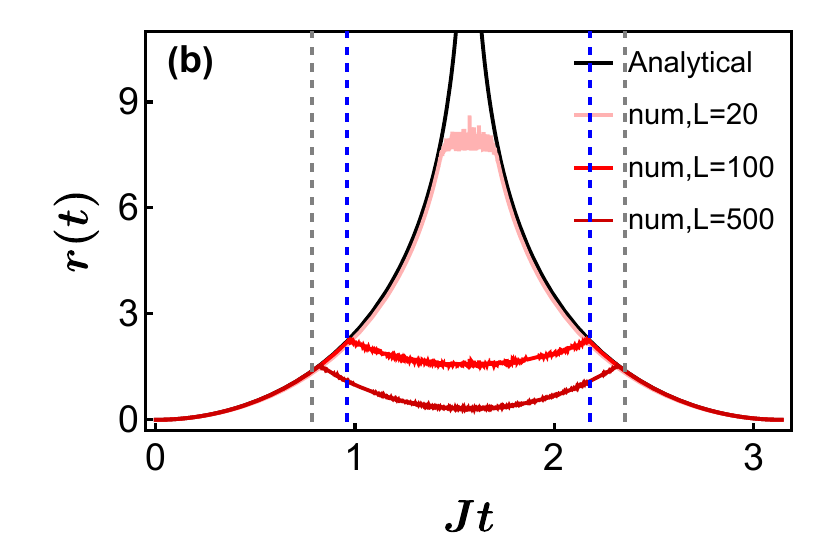}
    \caption{(a) Analytical LRFs under PBC and OBC. (b) LRFs under OBC computed from the analytical expression \eqref{eq:r_OBC} and the numerical evaluation using Eq. \eqref{eq:r_num}. The red curves in panel (b) show numerical results for different system sizes; from lighter to darker red, the system size increases. The two gray dashed vertical lines are located at $Jt=\pi/4$ and $Jt=3\pi/4$, respectively. In the interval between these two lines ($\pi/4<Jt<3\pi/4$), LRFs for PBC and OBC differ. The two blue dashed vertical lines mark the points where the numerical result for $L=100$ begins to deviate visibly from the analytical curve. }
    \label{fig:PBC_OBC_diff}
\end{figure}

Comparing Eqs. \eqref{eq:r_PBC} and \eqref{eq:r_OBC}, we find that the thermodynamic LRF is boundary-dependent. The PBC and OBC results coincide only when $|\cos(Jt)|\ge|\sin(Jt)|$, or equivalently $-\pi/4+n\pi\le Jt\le\pi/4+n\pi,~n\in\mathbb{Z}$. Outside this interval, the two boundary conditions give distinct LRFs, as shown in Fig. \ref{fig:PBC_OBC_diff}(a). This provides an exact example in which the boundary condition affects the LRF in the thermodynamic limit, rather than only producing a finite-size correction, complementing the exact results previously obtained for other models \cite{PhysRevLett.123.160603,sbdw-2ly3}. This exact OBC result also provides a useful benchmark for testing numerical calculations. In particular, a reliable numerical method should reproduce the OBC result even in the boundary-dependent interval where it differs from the PBC result.

We therefore evaluate the LRF under OBC directly from the Slater determinant,
\begin{align}
    r^{\text{num}}_L(t)=-\frac{1}{L}\ln\left|\det{M_L(t)}\right|^2,
\end{align}
using double-precision. For a global quench, the many-body Loschmidt amplitude decreases exponentially with system size. Thus $|\det M_L(t)|$ rapidly approaches zero, and taking its logarithm directly is numerically unstable. In practice we compute the singular values $\sigma_\alpha(L,t)$ of $M_L(t)$ and use the equivalent expression
\begin{align}
    r^{\text{num}}_L(t)=-\frac{2}{L}\sum_{\alpha=1}^L\ln{\sigma_\alpha(L,t)}.
    \label{eq:r_num}
\end{align}
This avoids taking the logarithm of an exponentially small determinant. However, it does not solve the OBC instability: in the interval where the PBC and OBC LRFs differ, some singular values themselves become extremely small and eventually reach the precision limit, as will be discussed in the next section.

Figure \ref{fig:PBC_OBC_diff}(b) shows the first numerical symptom of the problem. The analytical OBC result is compared with the double-precision singular value evaluation for several system sizes. The numerical curves are reliable in the interval where the analytical LRFs under PBC and OBC coincide. In the boundary-dependent interval where the two analytical LRFs differ, the behavior is strongly size-dependent. For relatively small systems, the numerical curve can still agree with the analytical OBC result over a large part of this interval. As the system size increases, however, the numerical curve begins to deviate from the OBC benchmark. This does not mean that the true large-system OBC LRF deviates from the analytical expression. Instead, it shows that increasing the system size at fixed precision can make the numerical estimate less accurate.

\section{Singular value diagnosis of the boundary dependence and numerical instability}
\label{sec:singular_values}

\begin{figure}
    \centering
    \includegraphics[width=0.8\linewidth]{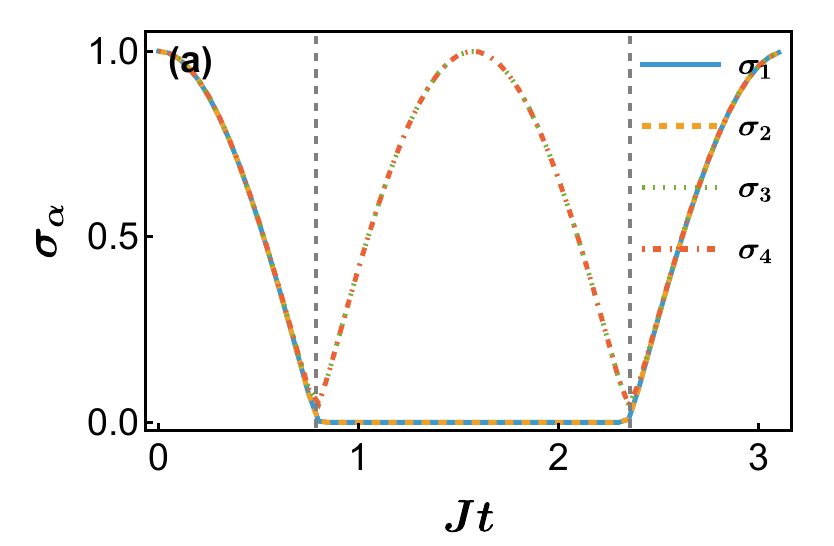}
    \includegraphics[width=0.8\linewidth]{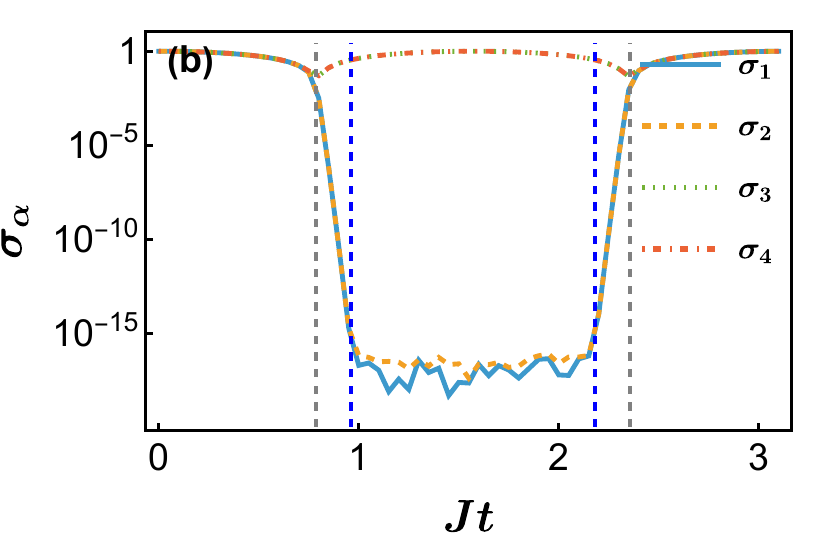}
    \includegraphics[width=0.8\linewidth]{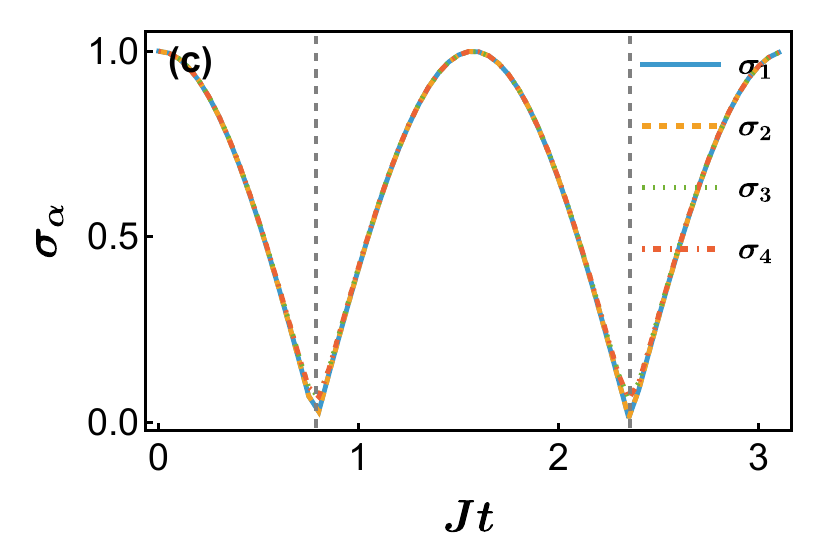}
    \caption{(a) The four smallest singular values of the OBC Loschmidt matrix for $L=100$, computed in double-precision. Panel (b) shows the same data as in (a) on a logarithmic scale. (c) The four smallest singular values of the PBC Loschmidt matrix for $L=100$. The gray and blue dashed vertical lines are the same as those in Fig. \ref{fig:PBC_OBC_diff}(b).}
    \label{fig:sVals}
\end{figure}

In this section, we analyze the singular value spectrum of the Loschmidt matrix in Eq. \eqref{eq:Loschmidt_mat}, and relate it to the boundary dependence and numerical instability of LRFs. It has been reported in Ref. \cite{PhysRevB.97.064304} that for the open-boundary SSH model, two eigenvalues of the Loschmidt matrix can approach zero during quenches from the topologically trivial phase to the topological phase, indicating DQPTs in the bulk \cite{PhysRevB.108.094306,Masłowski_2024,Masłowski_2026}. Here we focus instead on singular values, because the Loschmidt matrix is generally non-Hermitian and its singular values are more stable numerically.

Let the singular values of $M_L(t)$ be ordered as
\begin{align}
    \sigma_1(L,t)\le\sigma_2(L,t)\le\cdots\le\sigma_L(L,t).
\end{align}
In Fig. \ref{fig:sVals}(a) we show the four smallest singular values $\sigma_\alpha(L,t)~(\alpha=1,2,3,4)$ of the OBC Loschmidt matrix \eqref{eq:Loschmidt_mat} for $L=100$. The two gray dashed lines mark the interval in which the LRFs under PBC and OBC differ. Within this interval, two singular values become close to zero, while all other singular values remain finite. This correspondence indicates that the emergence of these near-zero singular values provides a direct diagnostic of the boundary-dependent interval. As we will show below, these near-zero singular values decay exponentially with system size and eventually vanish in the thermodynamic limit, and therefore qualify as TZSVs; the corresponding singular vectors are the DZMs. In Fig. \ref{fig:sVals}(b) we show the same data on a logarithmic scale. When the numerical LRF begins to deviate from the analytical result (indicated by the two blue dashed vertical lines), the singular values of DZMs are approximately $\mathcal{O}\left(10^{-16}\right)$, which is below the reliable range of double-precision. In the interval between the two blue dashed vertical lines, the evaluation of these singular values is no longer accurate, leading to the deviation of the numerical result from the analytical expression.

For comparison, we also plot the singular value spectrum of the PBC Loschmidt matrix (Eq. \eqref{eq:Loschmidt_mat_PBC}) in Fig. \ref{fig:sVals}(c). Unlike the OBC case, no singular values approach zero in the same time interval. This contrast shows that the singular value spectrum of the Loschmidt matrix depends on the boundary condition. More importantly, it indicates that the difference between LRFs under PBC and OBC originates from the DZMs of the OBC Loschmidt matrix.

\begin{figure}
    \centering
    \includegraphics[width=0.8\linewidth]{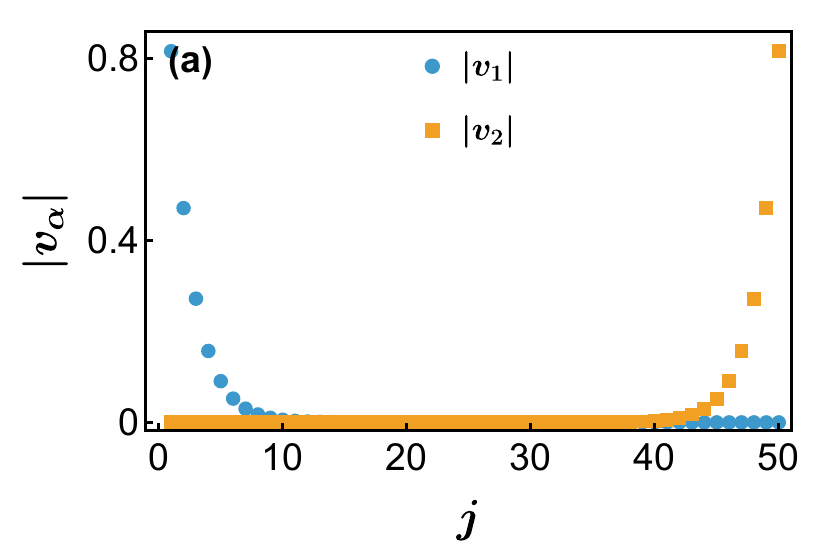}
    \includegraphics[width=0.8\linewidth]{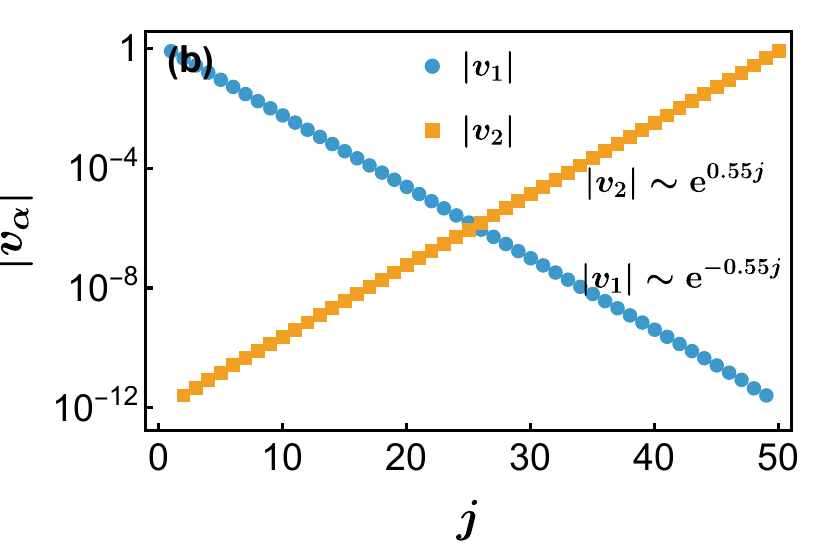}
    \caption{Two DZMs of $M_L(t)$ defined in Eq. \eqref{eq:Loschmidt_mat}, with $L=50$ and $Jt=\pi/3$. Panel (b) shows the same data on a logarithmic scale. The fitted decay agrees with the inverse localization length $\xi^{-1}=-\ln(\cot{\pi/3})\approx0.55$. }
    \label{fig:sVecs}
\end{figure}

We next show that the two smallest singular values vanish in the thermodynamic limit throughout the interval $|\cos(Jt)|<|\sin(Jt)|$, and hence the corresponding singular vectors are DZMs. To determine the condition for such TZSVs, we consider the corresponding zero-singular-value equation. For a matrix $M$ with a zero singular value and corresponding right singular vector $v_\alpha$, one has
\begin{align}
    Mv_\alpha=0.
\end{align}
For our Loschmidt matrix in Eq. \eqref{eq:Loschmidt_mat}, the bulk equation is
\begin{align}
    bv_{\alpha,j-1}+cv_{\alpha,j}+bv_{\alpha,j+1}=0,
    \label{eq:bulk_equation}
\end{align}
and the boundary equations are
\begin{align}
    av_{\alpha,1}+bv_{\alpha,2}&=0,\nn\\
    bv_{\alpha,L-1}+av_{\alpha,L}&=0.
    \label{eq:boundary_equations}
\end{align}
Making the ansatz $v_{\alpha,j}\sim z^j$ and substituting it into the bulk equation \eqref{eq:bulk_equation} and boundary equations \eqref{eq:boundary_equations}, we obtain
\begin{subequations}\begin{align}
    \frac{b}{z}+c+bz&=0,\label{eq:bulk}\\
    az+bz^2&=0,\label{eq:left}\\
    bz^{L-1}+az^L&=0.\label{eq:right}
\end{align}\end{subequations}
A right singular vector can be either extended, with $|z|=1$, or localized near an edge, with $|z|\ne1$. Here we numerically solve for the singular vectors associated with the TZSVs, and find they are localized near two boundaries, as shown in Fig. \ref{fig:sVecs}. For simplicity, we consider the edge mode at the left boundary, i.e., $|z|<1$. Then we only need to consider the bulk equation \eqref{eq:bulk} and left boundary equation \eqref{eq:left}. Solving these two equations, we obtain
\begin{subequations}\begin{align}
    &|z|=\left|\frac{a}{b}\right|<1,\label{eq:cond1}\\
    &a^2+b^2-ac=0.\label{eq:cond2}
\end{align}\end{subequations}
Using the definitions of $a,~b$, and $c$ in Eq. \eqref{eq:abc}, the second condition Eq. \eqref{eq:cond2} is identically satisfied. Equation \eqref{eq:cond1} therefore gives the condition for the existence of TZSVs
\begin{align}
    |\cos(Jt)|<|\sin(Jt)|.
\end{align}
Thus DZMs appear exactly in the interval where LRFs under PBC and OBC differ. This is the central diagnostic result: the boundary-dependent interval of the LRF can be identified from the singular value spectrum of the OBC Loschmidt matrix.

From Eq. \eqref{eq:cond1}, the localization length of the edge mode can be obtained
\begin{align}
    \xi=\left(-\ln\left|\frac{a}{b}\right|\right)^{-1}=\left(-\ln|\cot(Jt)|\right)^{-1},
\end{align}
which is verified numerically in Fig. \ref{fig:sVecs}(b).

\begin{figure}
    \centering
    \includegraphics[width=0.8\linewidth]{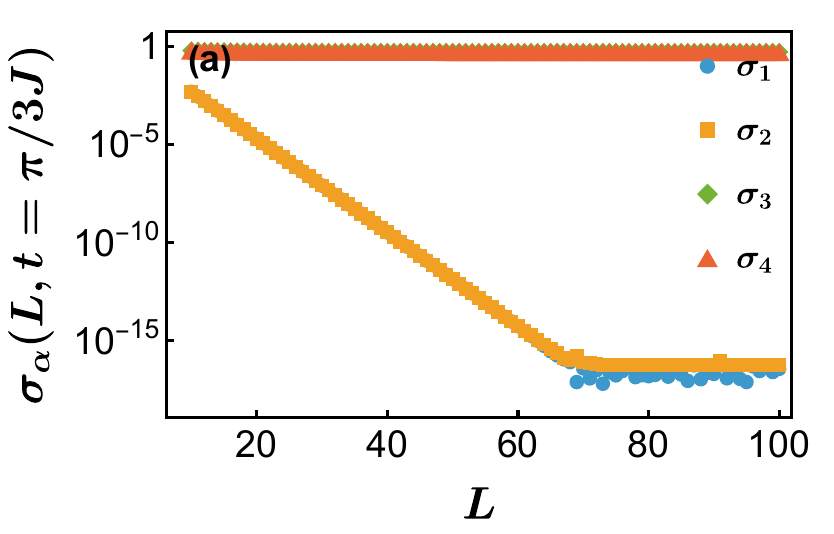}
    \includegraphics[width=0.8\linewidth]{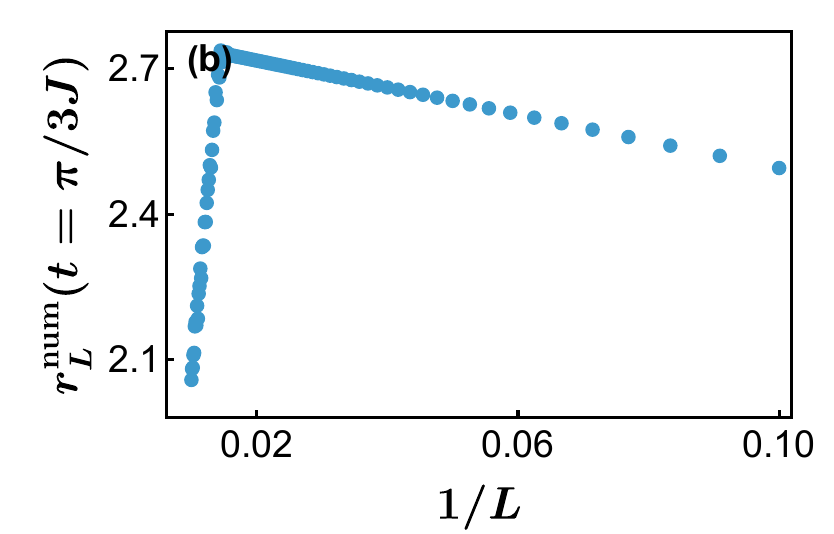}
    \caption{Finite-size scaling at $Jt=\pi/3$. (a) The four smallest singular values vs $L$. The two TZSVs $\sigma_{1,2}$ decay exponentially with system size until they reach the double-precision floor near $\mathcal{O}\left(10^{-16}\right)$, whereas $\sigma_{3,4}$ remain finite. (b) The LRF vs $1/L$, showing the reliable small-size branch and the precision-affected large-size branch. }
    \label{fig:FSS}
\end{figure}

Now we investigate the finite-size scaling of both the singular values and the LRF under OBC. We fix $Jt=\pi/3$, a time at which DZMs emerge, and compute these quantities in double-precision for different system sizes. In Fig. \ref{fig:FSS}(a), we show that for small system sizes, the two TZSVs $\sigma_{1,2}$ decay exponentially with $L$. Around $L\simeq69$, they reach approximately $\mathcal{O}\left(10^{-16}\right)$, at which point they can no longer be resolved accurately in double-precision. In contrast, the larger singular values $\sigma_{3,4}$ remain finite with increasing $L$. We emphasize that the loss of accuracy for the TZSVs does not imply an instability of the singular-value decomposition (SVD) algorithm itself. Standard SVD algorithms are stable: the error of the singular values of a matrix $M$ is bounded by $p(n)\epsilon||M||_2$, where $||\cdot||_2$ is the 2-norm of a matrix, $\epsilon$ is the machine precision, and $p(n)$ is a slowly growing function of the dimension $n$ of $M$ \cite{doi:10.1137/0911052,doi:10.1137/1.9781421407944}. Importantly, this is an absolute rather than a relative error bound, so sufficiently small singular values may lose relative accuracy even though their absolute errors remain small. For a TZSV scaling as $\sigma(L)\simeq C\ee^{-\kappa L}$, numerical accuracy is therefore expected to deteriorate once $\sigma(L)$ approaches the finite-precision error scale. In our double-precision calculations, this occurs when the TZSVs reach values of order $\mathcal{O}\left(10^{-16}\right)$, which determines the crossover size beyond which the numerical results depart from the reliable small-size branch shown in Fig. \ref{fig:FSS}. In practice, the reliable finite-size regime can be identified from the range in which the smallest singular values follow their expected scaling behavior and, when necessary, verified by comparing calculations performed at different numerical precisions.

In Fig. \ref{fig:FSS}(b) we show the corresponding finite-size scaling of the LRF. In double-precision, the LRF appears to be piecewise linear as a function of $1/L$. We emphasize that only the small-size branch is reliable. Therefore, linear fitting of this branch provides a practical way to estimate the LRF in the thermodynamic limit under OBC even when larger systems are affected by finite-precision errors. For example, fitting the reliable small-size data gives
\begin{align}
    r_L^{\text{num}}\left(\frac{\pi}{3J}\right)=2.77-\frac{2.77}{L},
    \label{eq:linear_fitting}
\end{align}
which produces the analytical value $r_{\text{OBC}}(\pi/(3J))=-2\ln\cos^2(\pi/3)\approx2.77$ when $L\to\infty$. The turning point between the two branches occurs near $1/L\approx0.0145$, or $L\simeq69$, the same size at which the two TZSVs reach the double-precision floor. This agreement shows that the large-size branch is a finite-precision artifact, while the reliable small-size branch can be used as a feasible extrapolation method for the LRF in the thermodynamic limit under OBC.

\begin{figure}
    \centering
    \includegraphics[width=0.8\linewidth]{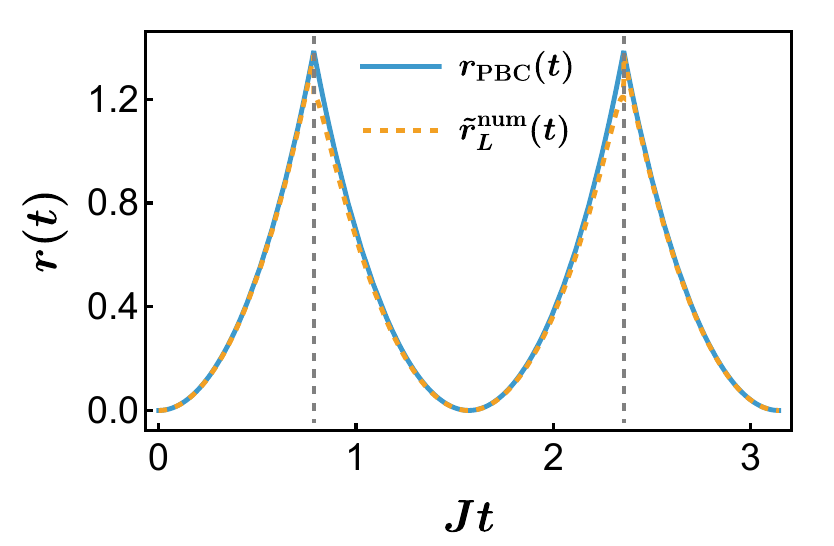}
    \caption{Comparison between the LRF under PBC and the modified LRF under OBC, for $L=100$. The gray dashed vertical lines are the same as those in Fig. \ref{fig:PBC_OBC_diff}(b). The agreement shows that the difference between LRFs under PBC and OBC is carried by the two DZMs.}
    \label{fig:tilde_r}
\end{figure}

We now show that the two DZMs carry the boundary-dependent part of the LRF. We first emphasize that the contribution of DZMs does not vanish in the thermodynamic limit. The reason is that the LRF depends on the logarithm of singular values. For a DZM, the associated finite-size singular value decays exponentially with system size, $\sigma_\alpha(L,t)\sim e^{-\kappa_\alpha L}$. Consequently, $\ln\sigma_\alpha\sim-\kappa_\alpha L$, and the factor $1/L$ in the definition of the LRF converts this exponentially small singular value into a finite contribution. To separate this contribution, we define a modified LRF under OBC by omitting the singular values of the two DZMs inside the interval where they become near zero
\begin{align}
    \tilde{r}_{L}^{\text{num}}(t)=
    \begin{cases}
        -\frac{2}{L}\sum_{\alpha=3}^L{\ln{\sigma_{\alpha}(L,t)}}, & \text{when DZMs emerge},\\
        -\frac{2}{L}\sum_{\alpha=1}^L{\ln{\sigma_{\alpha}(L,t)}}, & \text{without DZMs}.
    \end{cases}
    \label{eq:r_tilde}
\end{align}
As shown in Fig. \ref{fig:tilde_r}, this modified LRF under OBC agrees well with the analytical LRF under PBC. Therefore, after the contribution of the two DZMs is removed, the remaining singular value spectrum reproduces the PBC result. This demonstrates that the two DZMs encode the entire boundary-dependent difference between LRFs under PBC and OBC.

\section{General quench protocols in topological free-fermion models}
\label{sec:other}

\subsection{SSH model}
In this section, we examine whether the connection among DZMs, boundary dependence of the LRF, and finite-precision instability persists in other quench protocols. We first consider a general quench for the SSH model from the topologically trivial phase to the topological phase, i.e., $-1<\delta_i<0$ and $0<\delta_f<1$. As a representative example, we choose $\delta_i=-0.6$ and $\delta_f=0.6$. In this case, an analytical expression for the LRF under OBC is not available. Therefore, the singular value spectrum provides an important diagnostic for identifying whether the LRF depends on the boundary condition.

\begin{figure}
    \centering
    \includegraphics[width=0.8\linewidth]{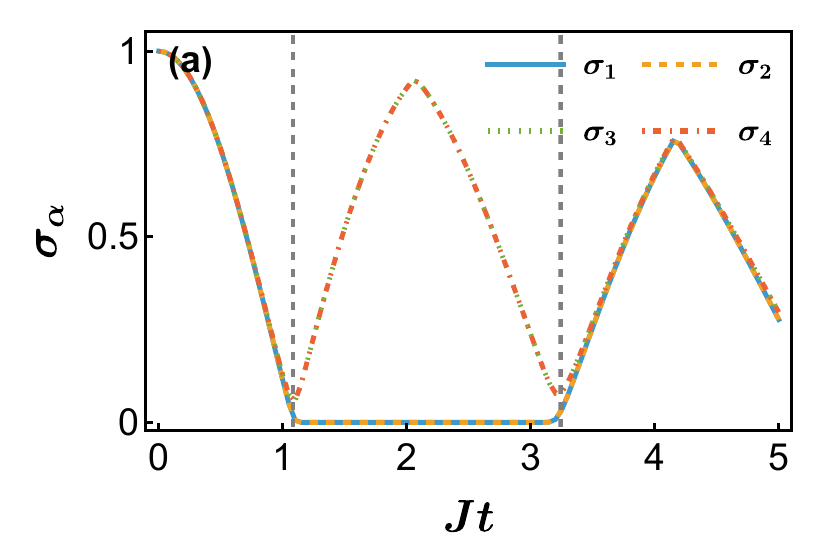}
    \includegraphics[width=0.8\linewidth]{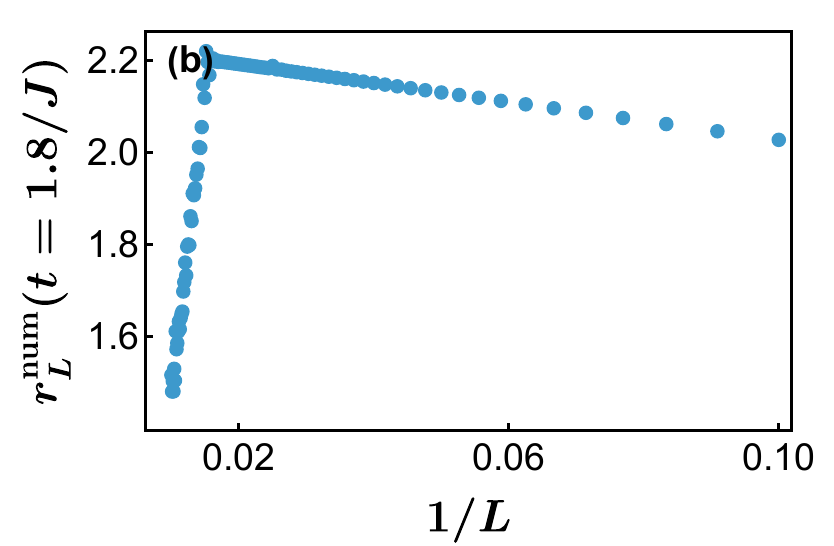}
    \includegraphics[width=0.8\linewidth]{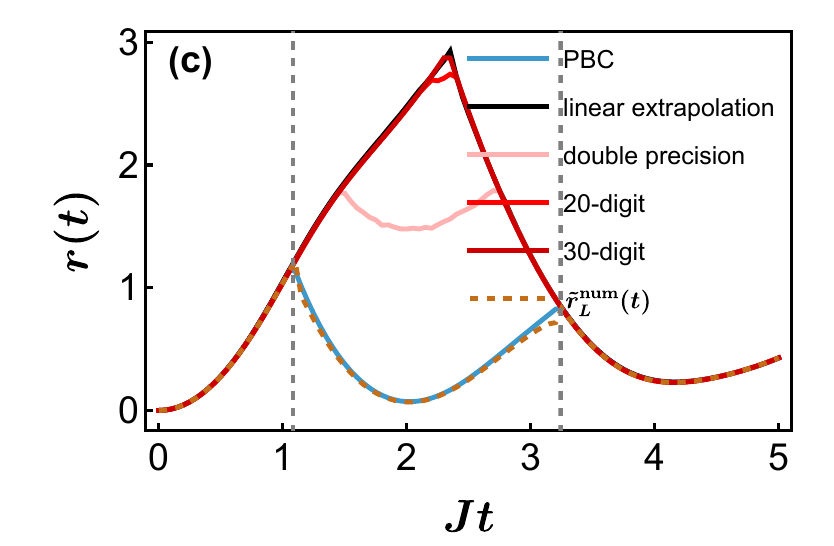}
    \caption{General quench from the topologically trivial phase to the topological phase in the SSH model, with $\delta_i=-0.6$ and $\delta_f=0.6$. (a) The four smallest singular values of the Loschmidt matrix for $L=100$. (b) Finite-size scaling of the LRF at $Jt=1.8$. (c) LRFs under PBC (blue curve) and OBC (red and black curves), and modified LRF under OBC (brown dashed curve). The red curves are computed with different precisions; from lighter to darker red, the precision increases. The black curve is obtained from linear extrapolation of the reliable small-size branch. The brown dashed curve removes the contribution of the two DZMs. The gray dashed vertical lines in panels (a) and (c) mark the time interval in which the LRFs under PBC and OBC differ. }
    \label{fig:trivial_to_topo}
\end{figure}

In Fig. \ref{fig:trivial_to_topo}(a) we show the four smallest singular values of the Loschmidt matrix under OBC. A pair of singular values becomes much smaller than the rest in a finite time interval, indicating the emergence of two DZMs. The two gray dashed vertical lines indicate the interval in which the LRFs under PBC and OBC differ, as extracted from Fig. \ref{fig:trivial_to_topo}(c). The coincidence between this interval and the interval where the two smallest singular values become near zero demonstrates that the DZMs diagnose the boundary-dependent part of the LRF.


The numerical consequence of these DZMs is shown in Fig. \ref{fig:trivial_to_topo}(b). At a representative time $Jt=1.8$, which lies inside the interval where the two DZMs emerge, the finite-size scaling of the LRF under OBC again exhibits two apparent linear branches as a function of $1/L$. The small-size branch corresponds to the regime in which the TZSVs are still resolved in double-precision. The large-size branch appears only after these singular values approach the precision floor and is therefore a finite-precision artifact. As in the exactly solvable case, the reliable small-size branch provides a practical way to estimate the LRF in the thermodynamic limit under OBC by linear extrapolation.

We apply this linear-extrapolation procedure throughout the relevant time interval and obtain the LRF under OBC shown by the black curve in Fig. \ref{fig:trivial_to_topo}(c). To verify this estimate, we also compute the LRF directly from Eq. \eqref{eq:r_num} with higher precisions. As the precision is increased, the direct numerical result converges to the extrapolated curve. In particular, the 30-digit result agrees well with the small-size extrapolation. This confirms that the extrapolated curve gives the correct LRF under OBC, while the double-precision deviation is caused by the loss of accuracy in the TZSVs.

In Fig. \ref{fig:trivial_to_topo}(c), we also compare the LRF under PBC with the LRF under OBC and with the modified LRF under OBC defined by removing the contribution of the two DZMs (see Eq. \eqref{eq:r_tilde}). The two gray dashed vertical lines mark the interval where LRFs under PBC and OBC differ. This interval agrees well with the interval in Fig. \ref{fig:trivial_to_topo}(a) where the two DZMs emerge. The modified LRF under OBC agrees well with the PBC result, indicating that the difference between LRFs under PBC and OBC is carried by the two DZMs. Therefore, for this more general quench, the two DZMs diagnose the interval where the LRF is boundary dependent, and their contribution accounts for the difference between LRFs under PBC and OBC.

\begin{figure*}
    \centering
    \includegraphics[width=0.3\linewidth]{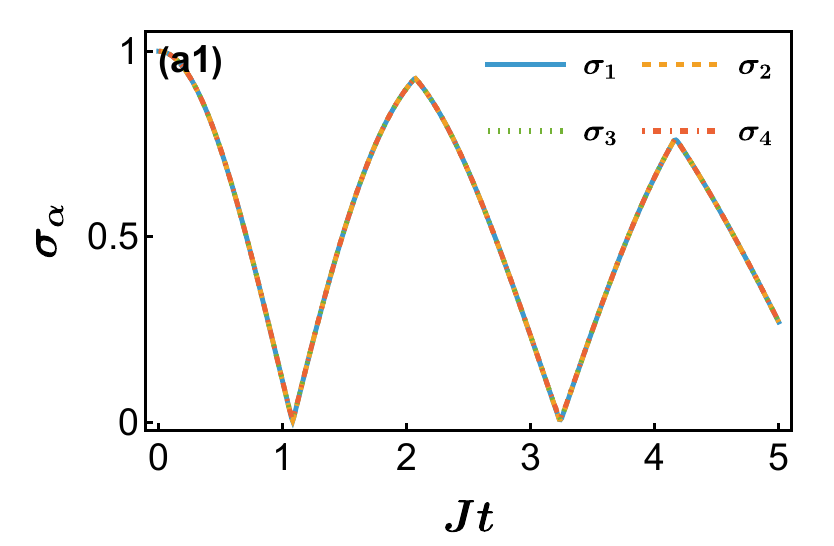}
    \includegraphics[width=0.3\linewidth]{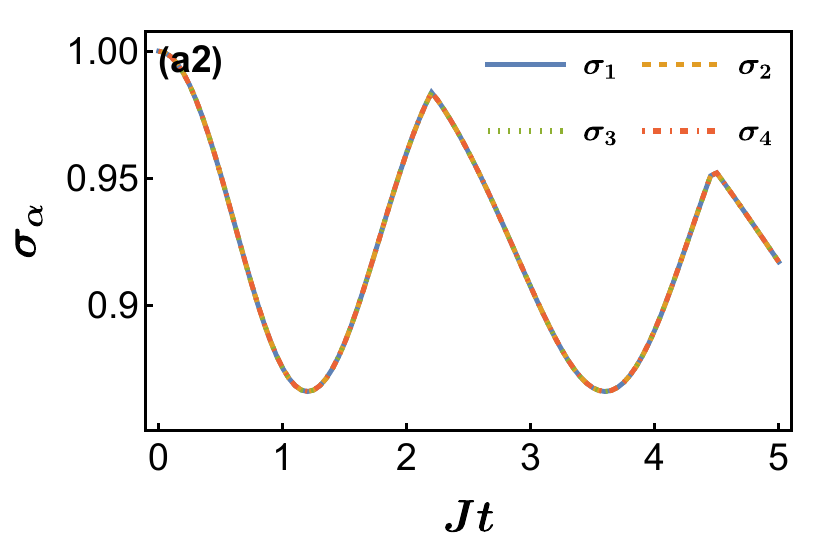}
    \includegraphics[width=0.3\linewidth]{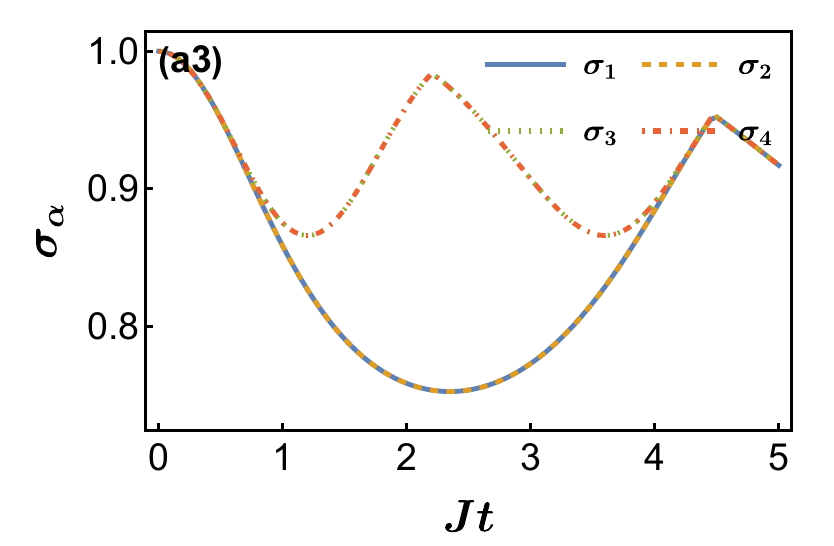}
    \includegraphics[width=0.3\linewidth]{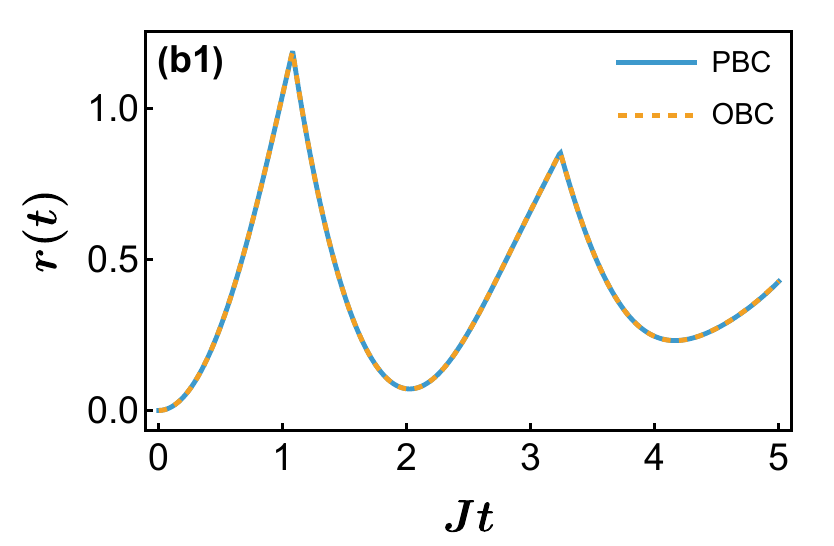}
    \includegraphics[width=0.3\linewidth]{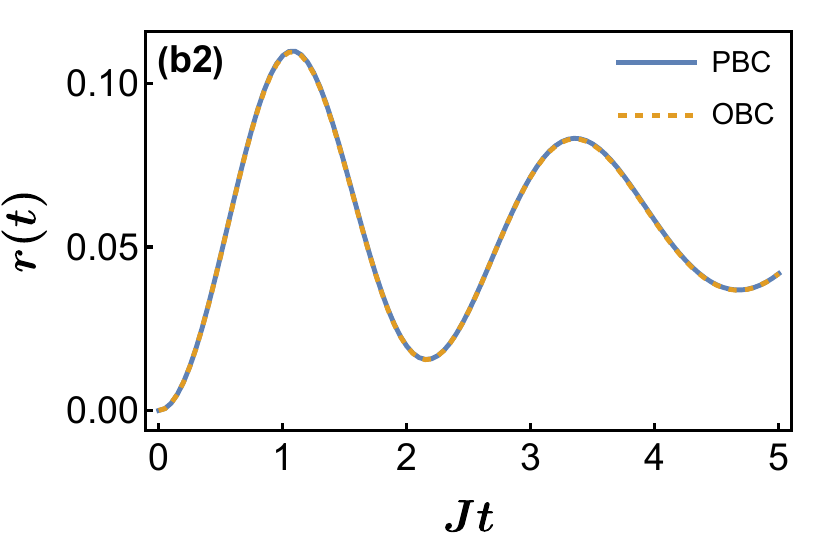}
    \includegraphics[width=0.3\linewidth]{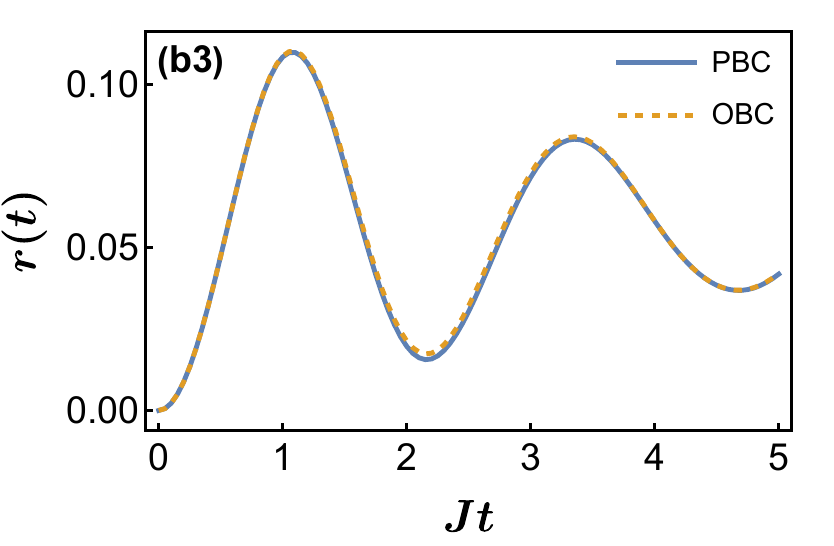}
    \caption{(a1)--(a3) The four smallest singular values and (b1)--(b3) LRFs for other quench protocols in the SSH model. (a1) and (b1) From the topological phase to the topologically trivial phase, with $\delta_i=0.6,~\delta_f=-0.6$; (a2) and (b2) from the topologically trivial phase to the topologically trivial phase, with $\delta_i=-0.2,~\delta_f=-0.6$; (a3) and (b3) from the topological phase to the topological phase, with $\delta_i=0.2,~\delta_f=0.6$. The numerical results for OBC are computed with the system size $L=600$. }
    \label{fig:other}
\end{figure*}

We also consider quenches from the topological phase to the topologically trivial phase, from the topologically trivial phase to the topologically trivial phase, and from the topological phase to the topological phase. In contrast to the quenches discussed earlier, no singular values become vanishingly small over a finite time interval, as shown in Figs. \ref{fig:other}(a1)--(a3). Correspondingly, Figs. \ref{fig:other}(b1)--(b3) suggest that LRFs under PBC and OBC coincide for these protocols. This supports the conclusion that the difference between LRFs under PBC and OBC is tied to the emergence of DZMs.

\subsection{Extended SSH model}
To further demonstrate the generality of our findings, we consider the extended SSH model, the Hamiltonian of which is given by \cite{PhysRevB.89.085111,Maffei_2018,PhysRevB.110.054312}
\begin{align}
    H=\sum_j&\left(t_ac_{j,A}^\dagger c_{j,B}+t_bc_{j,B}^\dagger c_{j+1,A}\right.\nn\\
    &\left.+t_cc_{j,A}^\dagger c_{j+1,B}+t_dc_{j,B}^\dagger c_{j+2,A}+\text{H.c.}\right),
\end{align}
where $t_\alpha~(\alpha=a,b,c,d)$ are hopping amplitudes, $A$ and $B$ label two sublattices, and H.c. represents Hermitian conjugate. The number of unit cells is $L$. For simplicity, we fix $t_a=t_b=-1$. Under PBC, after performing the Fourier transform, the Hamiltonian can be cast into the standard two-band form of Eq. \eqref{eq:two-band} with
\begin{align}
    d_k^x&=-1+(-1+t_c)\cos{k}+t_d\cos(2k),\nn\\
    d_k^y&=(-1-t_c)\sin{k}+t_d\sin(2k),\nn\\
    d_k^z&=0.
\end{align}
The extended SSH model exhibits a rich equilibrium phase diagram, allowing us to investigate quench dynamics between different topological phases.

\begin{figure*}
    \centering
    \includegraphics[width=0.3\linewidth]{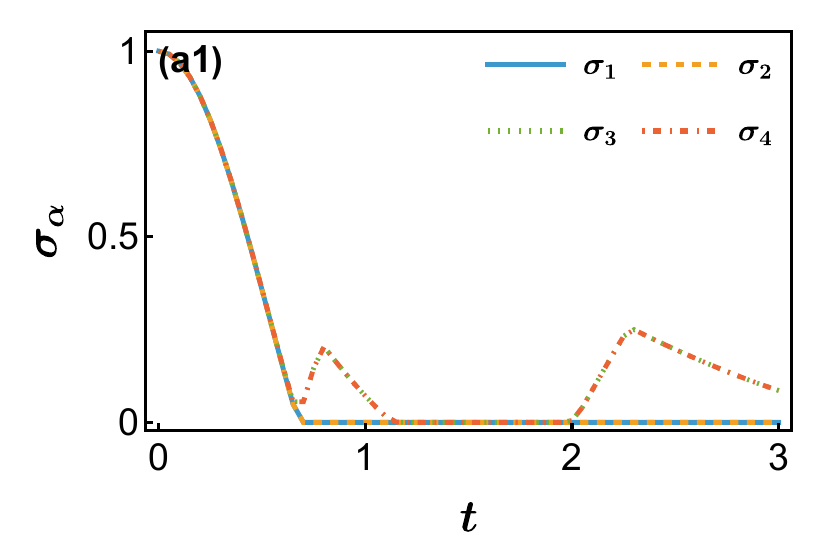}
    \includegraphics[width=0.3\linewidth]{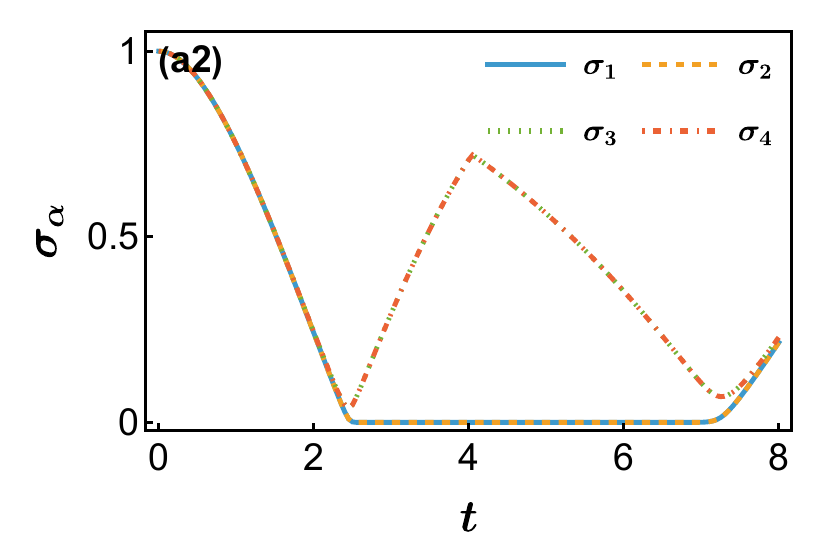}
    \includegraphics[width=0.3\linewidth]{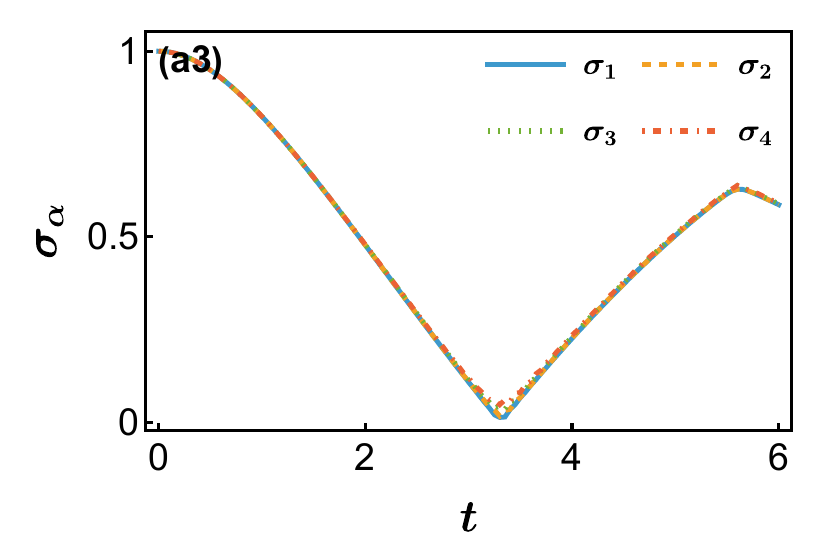}
    \includegraphics[width=0.3\linewidth]{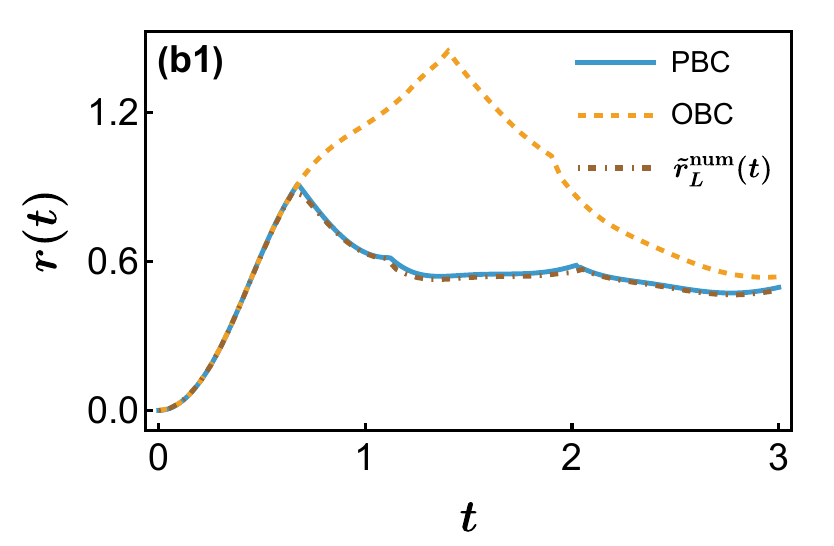}
    \includegraphics[width=0.3\linewidth]{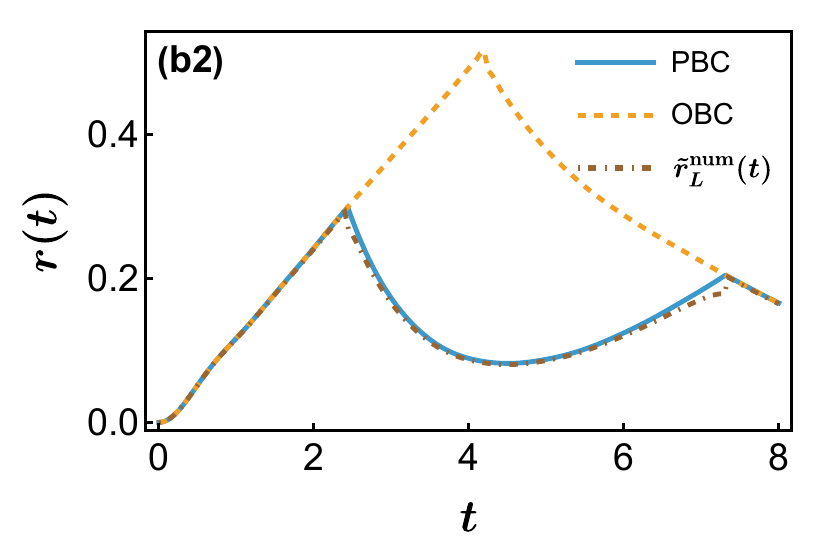}
    \includegraphics[width=0.3\linewidth]{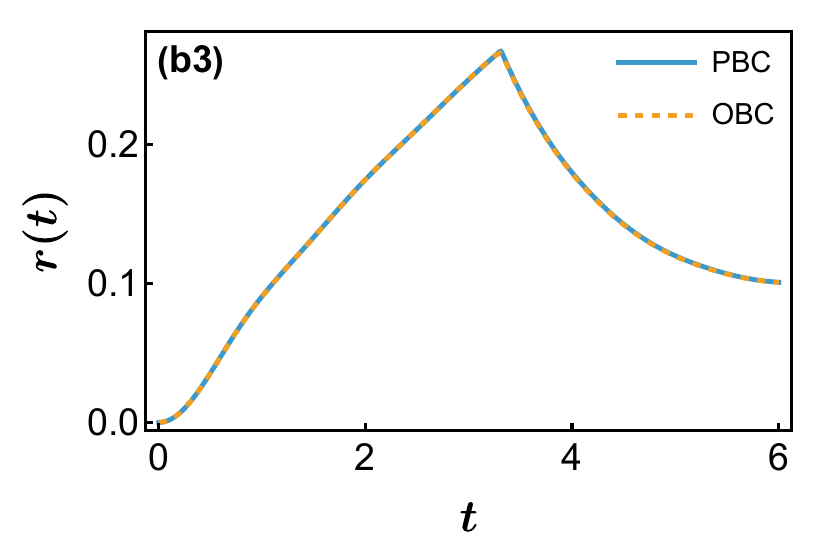}
    \caption{(a1)--(a3) The four smallest singular values and (b1)--(b3) LRFs for different quench protocols in the extended SSH model. (a1) and (b1) From $\nu=0$ to 2, with $t_{ci}=t_{cf}=0.5,~t_{di}=-0.5$, and $t_{df}=2$; (a2) and (b2) from $\nu=1$ to 2, with $t_{ci}=t_{cf}=0.5,~t_{di}=1.2$, and $t_{df}=2$; (a3) and (b3) from $\nu=2$ to 1, with $t_{ci}=t_{cf}=0.5,~t_{di}=2$, and $t_{df}=1.2$. The singular values are computed with the system size $L=600$. }
    \label{fig:ESSH}
\end{figure*}

We next examine the singular value spectrum of the Loschmidt matrix under OBC and the corresponding LRFs for quenches between different equilibrium phases of the extended SSH model. The winding number is denoted by $\nu$. Our numerical results show that DZMs appear for quenches that either increase the number of edge modes, such as $\nu=0\to\nu=\pm1,2$ and $\nu=\pm1\to\nu=2$, or reverse the winding orientation, such as $\nu=-1\to\nu=1$. Representative examples are shown in Figs. \ref{fig:ESSH}(a1)--(a2). For the quench from $\nu=0$ to $\nu=2$, up to four DZMs are found (Fig. \ref{fig:ESSH}(a1)), consistent with the increase in the total number of edge modes between the initial and final Hamiltonians. The corresponding LRFs under PBC and OBC are shown in Figs. \ref{fig:ESSH}(b1)--(b2). The thermodynamic OBC results are obtained by linear extrapolation from the reliable small-size branch, and we have checked that they agree well with the results computed directly from Eq. \eqref{eq:r_num} with a higher precision (not shown here). Whenever DZMs emerge, LRFs under PBC and OBC become different, demonstrating that DZMs diagnose the boundary dependence of LRFs. To isolate the contribution of these modes, we also compute the modified LRF under OBC by excluding all singular values identified as TZSVs. In particular, for the quench from $\nu=0$ to $\nu=2$, all four TZSVs are removed in the time interval where they appear. For all representative protocols, the modified LRF under OBC agrees well with the PBC result, confirming that the boundary-dependent contribution to the LRF is carried by the DZMs.

We also examine quenches in which the number of edge modes decreases, such as $\nu=2\to\nu=\pm1,0$ and $\nu=\pm1\to\nu=0$. For these protocols, no DZMs are observed. Correspondingly, the LRFs under PBC and OBC coincide. A representative example is shown in Figs. \ref{fig:ESSH}(a3) and (b3). This comparison further supports the conclusion that the boundary-dependent contribution to the LRF is tied to the emergence of DZMs.

The above results suggest a close connection between DZMs and equilibrium edge modes. In the increasing-edge-mode quenches, the maximum number of DZMs matches the increase in the total number of edge modes. For example, the $\nu=0\to\nu=2$ quench exhibits up to four DZMs, consistent with the appearance of two additional edge modes at each boundary. This correspondence is in line with the general idea that quench dynamics in one-dimensional two-band systems encodes the topological difference between the initial and final Hamiltonians \cite{PhysRevB.97.060304,PhysRevLett.121.250601}. 

\subsection{Kitaev chain}

To demonstrate that the DZM framework is not restricted to particle-number-conserving chiral systems, we further consider the Kitaev chain \cite{Kitaev_2001}, a paradigmatic model for topological superconductivity. Its Hamiltonian is
\begin{align}
    H=&-J\sum_j{\left(c_j^\dagger c_{j+1}+\text{H.c.}\right)}-\mu\sum_j{\left(c_j^\dagger c_j-\frac{1}{2}\right)}\nn\\
    &+\Delta\sum_{j}{\left(c_j c_{j+1}+\text{H.c.}\right)},
\end{align}
where $J$ is the hopping amplitude, $\Delta$ is the $p$-wave pairing amplitude, $\mu$ is the chemical potential, and H.c. represents Hermitian conjugate. For $\Delta\neq0$, the system is topologically nontrivial for $|\mu|<2|J|$ and topologically trivial for $|\mu|>2|J|$.

Using the Nambu basis $\Psi=(c_1,\cdots,c_L,c_1^\dagger,\cdots,c_L^\dagger)^T$,
the Hamiltonian under OBC can be written in terms of the $2L\times2L$ single-particle Bogoliubov-de Gennes (BdG) Hamiltonian $H_{\text{BdG}}$ as
\begin{align}
    H=\frac{1}{2}\Psi^\dagger H_{\mathrm{BdG}}\Psi,\qquad H_{\mathrm{BdG}}
    =
    \begin{pmatrix}
    A & B\\
    -B & -A
    \end{pmatrix},
\end{align}
where the nonzero matrix elements of the two $L\times L$ blocks $A$ and $B$ are
\begin{align}
    &A_{jj}=-\mu,\qquad A_{j,j+1}=A_{j+1,j}=-J,\nn\\
    &B_{j,j+1}=-B_{j+1,j}=-\Delta.
\end{align}
Within this BdG formalism, the Loschmidt echo can be conveniently obtained from a Slater determinant \cite{Masłowski_2026}
\begin{align}
    \left|\mathcal{G}(t)\right|^2=\left|\det{M(t)}\right|,\quad M_{ij}=\langle\phi_i|\ee^{-\ii H_{\text{BdG}}^ft}|\phi_j\rangle,
\end{align}
where $|\phi_i\rangle$ is the $i$th negative-energy eigenstate of the initial BdG Hamiltonian. In contrast to the particle-number-conserving case, the determinant here directly gives the Loschmidt echo $|\mathcal{G}(t)|^2$, rather than the Loschmidt amplitude $\mathcal{G}(t)$. Accordingly, the LRF is evaluated as $r_L^{\text{num}}(t)=-(1/L)\sum_{\alpha}\ln \sigma_\alpha(L,t)$ in terms of singular values of $M(t)$.

Under PBC, after performing the Fourier transform, the Hamiltonian can be written in the two-band form as
\begin{align}
    H=\sum_{k>0}{
    \begin{pmatrix}
        c_{k}^\dagger & c_{-k}
    \end{pmatrix}
    \boldsymbol{d}_k\cdot\boldsymbol{\sigma}
    \begin{pmatrix}
        c_{k} \\ c_{-k}^\dagger
    \end{pmatrix}},
\end{align}
with
\begin{align}
    \boldsymbol{d}_k=(0,2\Delta\sin{k},-2J\cos{k}-\mu).
\end{align}
The LRF can be directly evaluated from Eq. \eqref{eq:r_PBC1}, where, due to the restriction $k>0$, the momentum integration should be taken over $k\in[0,\pi]$.

\begin{figure*}
    \centering
    \includegraphics[width=0.3\linewidth]{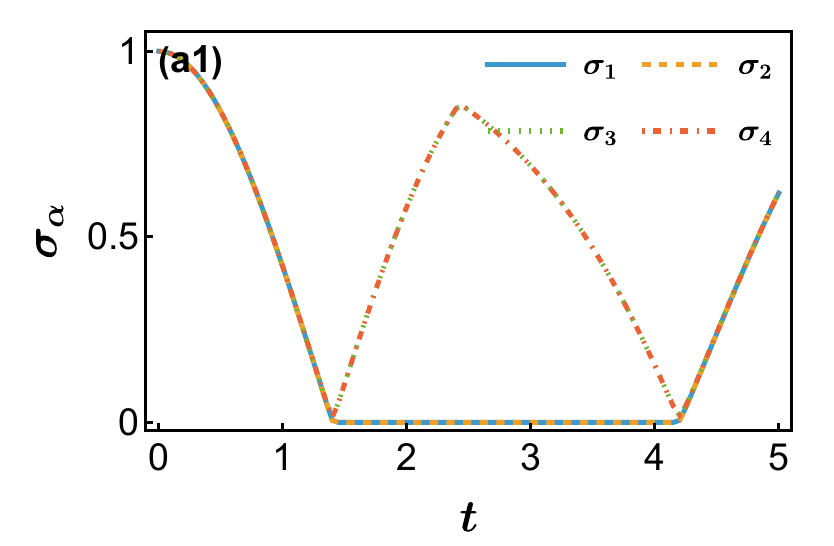}
    \includegraphics[width=0.3\linewidth]{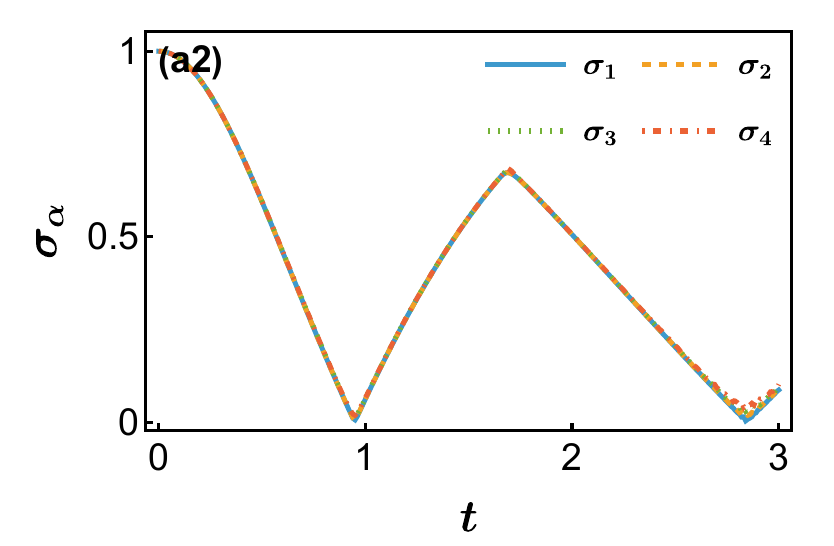}
    
    \includegraphics[width=0.3\linewidth]{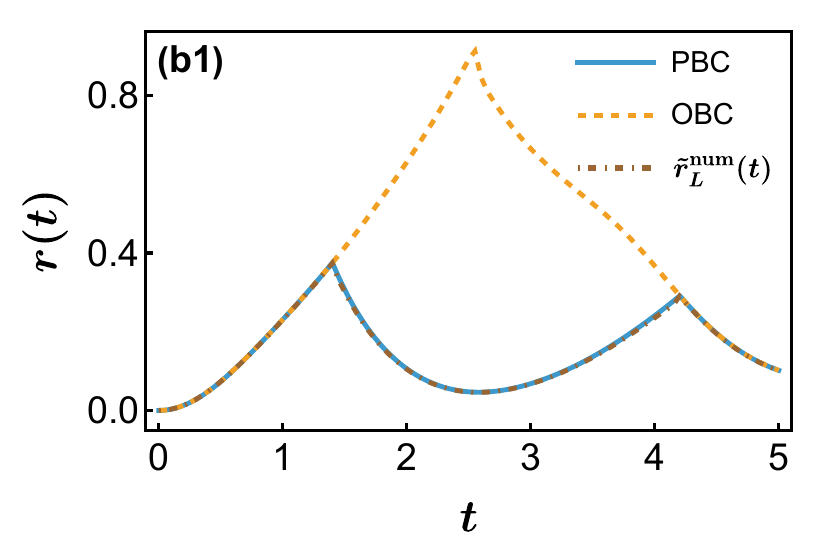}
    \includegraphics[width=0.3\linewidth]{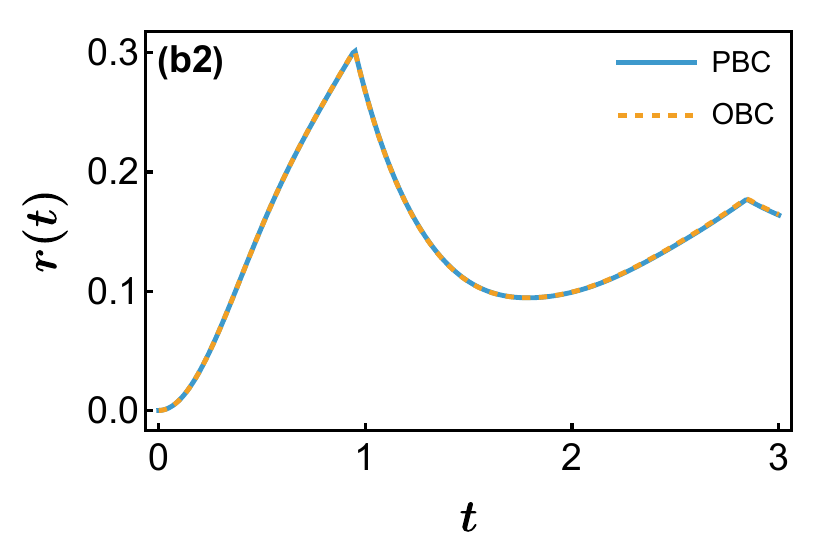}
    \caption{(a1), (a2) The four smallest singular values and (b1), (b2) LRFs for different quench protocols in the Kitaev chain. (a1) and (b1) From the topologically trivial phase to the topological phase, with $\mu_i=3,~\mu_f=1$; (a2) and (b2) from the topological phase to the topologically trivial phase, with $\mu_i=1,~\mu_f=3$. The singular values are computed with the system size $L=600$. }
    \label{fig:Kitaev}
\end{figure*}

We then examine the singular-value spectrum of the Loschmidt matrix under OBC and the corresponding LRFs. We set $J=1,~\Delta=0.8$ and first consider the quench from $\mu_i=3$ to $\mu_f=1$, i.e., from the topologically trivial phase to the topological phase. As shown in Fig. \ref{fig:Kitaev}(a1), two DZMs appear over a finite time interval, consistent with the appearance of one additional Majorana edge mode at each end of the final chain. The corresponding LRFs are shown in Fig. \ref{fig:Kitaev}(b1). The LRFs under PBC and OBC differ in the same time interval in which the two DZMs emerge. The OBC result is obtained by linear extrapolation from the reliable small-size branch. To isolate the contribution of the two DZMs, we also compute the modified LRF. The modified LRF under OBC agrees well with the PBC result, confirming that the boundary-dependent contribution is carried by the two DZMs.

For comparison, we also consider the reverse quench from $\mu_i=1$ to $\mu_f=3$, for which the number of Majorana edge modes decreases. No singular values become vanishingly small over a finite time interval, as shown in Fig. \ref{fig:Kitaev}(a2), and hence no DZMs emerge. Correspondingly, the LRFs under PBC and OBC coincide, as shown in Fig. \ref{fig:Kitaev}(b2). These results show that the connection among DZMs, boundary-dependent LRFs, and changes in equilibrium edge modes persists in the Kitaev chain, extending the DZM framework from particle-number-conserving chiral systems to topological superconductors described by BdG Hamiltonians.

\section{Summary and outlook}
\label{sec:conclusion}

In this work, we studied the role of DZMs in the boundary dependence and numerical evaluation of LRFs. Using the SSH chain as a representative free-fermion model, we found that LRFs under PBC and OBC differ in a finite time interval for quenches from the topologically trivial phase to the topological phase, and that this interval is precisely identified by the emergence of two DZMs in the OBC Loschmidt matrix. We further tested this correspondence in the extended SSH model and the Kitaev chain, which provide complementary examples with richer topological phases and superconducting pairing, respectively. For quenches that increase the number of edge modes or reverse the winding orientation, DZMs again emerge in the same time intervals where LRFs under PBC and OBC differ. 

The DZMs have two closely related consequences. On the physical side, they encode the boundary-dependent contribution to the LRF. This is confirmed by removing the contribution of DZMs from the LRF under OBC, which recovers the PBC result. On the numerical side, their finite-size singular values decay exponentially with system size and eventually become unresolved in fixed-precision arithmetic. This loss of resolution leads to an apparent deviation of the numerically computed LRF under OBC from the correct result. We showed that the reliable small-size branch, before the TZSVs reach the precision floor, can be used to estimate the LRF in the thermodynamic limit by linear extrapolation.

Our results also reveal a close connection between DZMs and equilibrium edge modes. In all the models studied in this work, the emergence and number of DZMs closely follow the change in the equilibrium edge-mode structure across the quench. This correspondence suggests that DZMs may be viewed as boundary-localized dynamical degrees of freedom associated with the topological mismatch between the initial and final Hamiltonians. Nevertheless, DZMs are singular vectors of the Loschmidt matrix rather than eigenstates of either Hamiltonian and therefore should not be identified directly with equilibrium edge modes. The observed correspondence is thus suggestive of a dynamical analog of bulk-boundary correspondence \cite{PhysRevLett.89.077002,RevModPhys.88.035005}, although establishing a general analytical formulation of this connection remains an open problem.

The present results also suggest several directions for future work. Since the mechanism identified here relies on the singular value spectrum of the Loschmidt matrix, it should be useful for analyzing boundary-dependent DQPTs in other free-fermion systems, including disordered, quasiperiodic, and non-Hermitian models. For disordered or quasiperiodic systems, localization and sample-to-sample fluctuations may complicate the identification of DZMs. For non-Hermitian systems, intrinsic boundary sensitivity, as manifested by the skin effect \cite{PhysRevLett.121.086803, PhysRevLett.124.086801,PhysRevLett.125.126402}, may constitute an additional source of boundary dependence in the LRFs and further aggravate numerical instability \cite{g1cw-tk7f}. A more fundamental challenge arises in interacting systems, where the Loschmidt amplitude is no longer reducible to the determinant of a single-particle Loschmidt matrix. Whether an appropriate many-body analog of DZMs can be defined, while retaining their connection to boundary dependence of LRFs, remains an open question. From a numerical perspective, our results indicate that the smallest singular values should be monitored carefully when evaluating LRFs in large systems. In particular, high-precision calculations or finite-size extrapolations in small-size systems may be necessary whenever exponentially small singular values appear.

\section*{Acknowledgments}
This work is supported by the NSFC under Grants No. 12474287, No. 12547107, and No. T2121001.

\section*{Data availability}
The data that support the findings of this article are not publicly available. The data are available from the authors upon reasonable request.

\appendix

\section{Derivation of the open-boundary Loschmidt matrix and its determinant}
\label{app1}

In this appendix, we derive Eq. \eqref{eq:Loschmidt_mat} and evaluate its determinant. For $\delta_f=1$, the final Hamiltonian consists of decoupled dimers on $(2,3),(4,5),\cdots,(2L-2,2L-1)$, together with two isolated edge sites $1$ and $2L$. On each dimer $(2j,2j+1)$, the one-particle Hamiltonian is
\begin{align}
    h_j
    =
    -2J
    \begin{pmatrix}
        0&1\\
        1&0
    \end{pmatrix}
    =
    -2J\sigma_x .
\end{align}
Hence
\begin{align}
    \ee^{-\ii h_j t}
    =
    \ee^{\ii 2Jt\sigma_x}
    =
    \cos(2Jt)I
    +
    \ii\sin(2Jt)\sigma_x .
\end{align}
Therefore
\begin{align}
    \ee^{-\ii H_f t}|2j\rangle
    &=
    \cos(2Jt)|2j\rangle
    +
    \ii \sin(2Jt)|2j+1\rangle,
    \nn\\
    \ee^{-\ii H_f t}|2j+1\rangle
    &=
    \ii \sin(2Jt)|2j\rangle
    +
    \cos(2Jt)|2j+1\rangle .\label{eq:evol}
\end{align}
The two isolated edge sites evolve trivially:
\begin{align}
    \ee^{-\ii H_f t}|1\rangle=|1\rangle,
    \qquad
    \ee^{-\ii H_f t}|2L\rangle=|2L\rangle .
\end{align}

We first consider a bulk occupied state $|\phi_j\rangle$ with $2\le j\le L-1$. Using Eq. \eqref{eq:evol}, we immediately obtain
\begin{align}
    \ee^{-\ii H_f t}|\phi_j\rangle=&\frac{1}{\sqrt2}\left(\ii\sin(2Jt)|2j-2\rangle+\cos(2Jt)|2j-1\rangle\right.\nn\\
    &\left.+\cos(2Jt)|2j\rangle+\ii\sin(2Jt)|2j+1\rangle\right).
\end{align}
Taking the overlap with $\langle \phi_i|=\left(\langle 2i-1|+\langle 2i|\right)/\sqrt{2}$, we obtain
\begin{align}
M_{ij}=\cos(2Jt)\delta_{i,j}+\frac{\ii\sin(2Jt)}{2}(\delta_{i-1,j}+\delta_{i+1,j}).
\end{align}
Thus the bulk part of the Loschmidt matrix is tridiagonal.

At the left boundary, $j=1$, the initial occupied orbital is $|\phi_1\rangle=\left(|1\rangle+|2\rangle\right)/\sqrt{2}$.
Since site $1$ is isolated while site $2$ belongs to the dimer $(2,3)$, this state evolves as
\begin{align}
    \ee^{-\ii H_f t}|\phi_1\rangle
    =
    \frac{1}{\sqrt{2}}
    \left(
        |1\rangle
        +
        \cos(2Jt)|2\rangle
        +
        \ii \sin(2Jt)|3\rangle
    \right).
\end{align}
Thus
\begin{align}
    M_{1,1}
    &=
    \frac{1+\cos(2Jt)}{2},
    \nn\\
    M_{2,1}
    &=
    \frac{\ii \sin(2Jt)}{2}.
\end{align}
Similarly, at the right boundary $j=L$, we have
\begin{align}
    M_{L,L}
    &=
    \frac{1+\cos(2Jt)}{2},
    \nn\\
    M_{L-1,L}
    &=
    \frac{\ii \sin(2Jt)}{2}.
\end{align}

Collecting these results, the Loschmidt matrix under OBC is
\begin{align}
    M_L(t)
    =
    \begin{pmatrix}
        a&b&0&0&\cdots&0\\
        b&c&b&0&\cdots&0\\
        0&b&c&b&\cdots&0\\
        0&0&b&c&\ddots&0\\
        \vdots&\vdots&\vdots&\ddots&\ddots&b\\
        0&0&0&0&b&a
    \end{pmatrix},
    \label{eq:Loschmidt_mat_app}
\end{align}
where
\begin{align}
    a=\frac{1+\cos(2Jt)}{2},
    \quad
    b=\frac{\ii\sin(2Jt)}{2},
    \quad
    c=\cos(2Jt).
    \label{eq:abc_app}
\end{align}
For the PBC case, the final Hamiltonian has an extra dimer on $(2L,1)$, and the corresponding Loschmidt matrix is
\begin{align}
    M_L^{\text{PBC}}(t)
    =
    \begin{pmatrix}
        c&b&0&0&\cdots&b\\
        b&c&b&0&\cdots&0\\
        0&b&c&b&\cdots&0\\
        0&0&b&c&\ddots&0\\
        \vdots&\vdots&\vdots&\ddots&\ddots&b\\
        b&0&0&0&b&c
    \end{pmatrix}.
    \label{eq:Loschmidt_mat_PBC}
\end{align}

We now evaluate the determinant of Eq. \eqref{eq:Loschmidt_mat_app}. Let $D_n$ denote the determinant of the leading $n\times n$ submatrix with the left boundary entry $a$ and bulk diagonal entries $c$, i.e.,
\begin{align}
    D_n
    =
    \det\begin{pmatrix}
        a&b&0&0&\cdots&0\\
        b&c&b&0&\cdots&0\\
        0&b&c&b&\cdots&0\\
        0&0&b&c&\ddots&0\\
        \vdots&\vdots&\vdots&\ddots&\ddots&b\\
        0&0&0&0&b&c
    \end{pmatrix}_{n\times n}.
\end{align}
Expanding the tridiagonal determinant along the last row and the last column, we obtain the recurrence relation
\begin{align}
D_n=cD_{n-1}-b^2D_{n-2},
\end{align}
with $D_1=a$ and $D_2=ac-b^2$. Solving this recurrence equation and using the definitions of $a, b$, and $c$ in Eq. \eqref{eq:abc_app}, we obtain an extremely simple expression
\begin{align}
    D_n=\left[\cos(Jt)\right]^{2n}.
\end{align}

Now we consider the determinant of the Loschmidt matrix $M_L(t)$. Expanding the determinant along the last row and the last column, we obtain
\begin{align}
    \det M_L(t)=aD_{L-1}-b^2D_{L-2}=\left[\cos(Jt)\right]^{2L-2},
\end{align}
which proves Eq. \eqref{eq:G_OBC}.

\bibliography{bibtex}

\end{document}